\documentclass[twocolumn]{revtex4-2}

\usepackage{amssymb}
\usepackage{amsmath}
\usepackage{dsfont}
\usepackage{mathtools}
\usepackage{graphicx}
\usepackage{subfig}
\usepackage{natbib}
\usepackage{hyperref}
\usepackage{url}
\usepackage{lipsum}
\usepackage{xcolor}

\begin{document}

\title{Learning of networked spreading models from noisy and incomplete data}

\author{Mateusz Wilinski}
\author{Andrey Y. Lokhov}
\affiliation{Theoretical Division, Los Alamos National Laboratory, Los Alamos, USA}

\begin{abstract}
    Recent years have seen a lot of progress in algorithms for learning parameters of spreading dynamics from both full and partial data. Some of the remaining challenges include model selection under the scenarios of unknown network structure, noisy data, missing observations in time, as well as an efficient incorporation of prior information to minimize the number of samples required for an accurate learning. Here, we introduce a universal learning method based on scalable dynamic message-passing technique that addresses these challenges often encountered in real data. The algorithm leverages available prior knowledge on the model and on the data, and reconstructs both network structure and parameters of a spreading model. We show that a linear computational complexity of the method with the key model parameters makes the algorithm scalable to large network instances.
\end{abstract}

\maketitle

\section{Introduction}

\begin{figure*}
    \includegraphics[width=\textwidth]{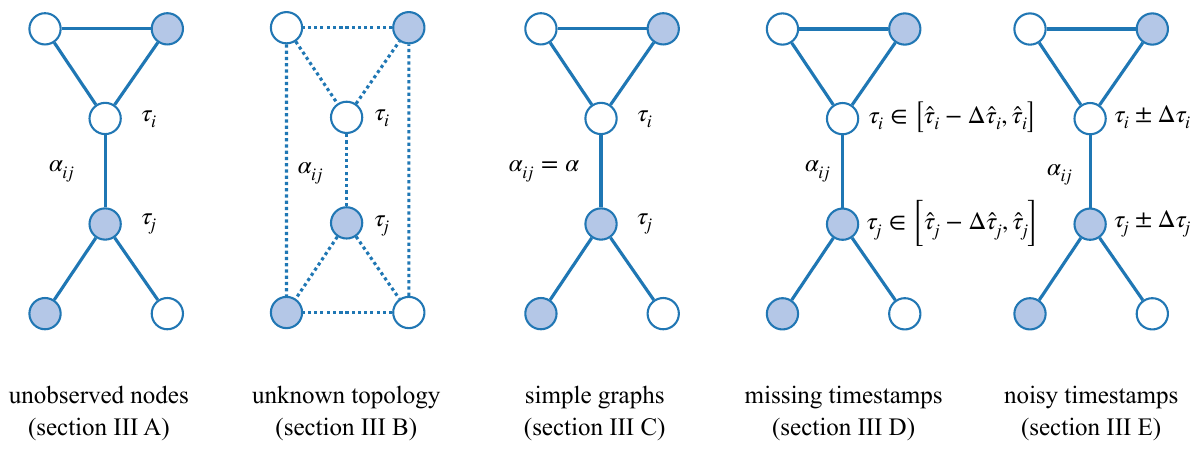}
    \caption{\label{fig:scenarios} Schematic representation of different scenarios of incomplete and uncertain data that we consider in learning of networked spreading models. We focus on the general setting where a fraction of the nodes never reports information (depicted as empty nodes). In addition, we treat cases where network structure is not known or known only partially; where the number of data samples is small, but prior information on the parameters is available; and where the observed node activation timestamps are missing or noisy. At the end of our experimental evaluation, we address learning of real-world network instances under a combination of these scenarios.}
\end{figure*}

Spreading models are routinely used to generate predictions for a plethora of diffusion processes on networks, whereby infectious diseases, opinions, or failures propagate in natural, social, and technological systems \cite{newman2018networks,barrat2008dynamical}. In these models, the nodes typically go from inactive to an active state through interactions with their neighbors on a network, similarly to how an infection is passed from one person to another. When the model structure or parameters are unknown, it is natural to consider the inverse problem of learning of the spreading model from data. Available data usually takes form of reported activation times for nodes in a network in several observed activation cascades. Whether due to a limited observation budget or imperfect reporting, the accessible data is unlikely to be perfect, and may be subject to uncertainty or provide only partial information on the system. This motivates the design of robust methods for selecting a spreading model from incomplete and noisy data. In this paper, we introduce a scalable learning method that addresses the challenges related to imperfect data.

Reconstruction of networked spreading models has been addressed by a number of works in recent years. In the context of full and exact observation of nodes' activation times, \cite{netrapalli2012learning, gomez2012inferring, abrahao2013trace, pouget2015inferring} showed that maximum-likelihood type approaches succeed to learn the model structure and parameters. The problem becomes significantly more difficult when the system is only partially observed, i.e., only a fraction of nodes report information about their activation times. In this case, the maximum likelihood approach has exponential complexity with respect to the number of unobserved nodes, which warrants learning methods based on the direct problem of predicting the model dynamics \cite{lokhov2016reconstructing}. This approach has lead to an efficient algorithm for learning the parameters of the spreading model on a known network, based on minimization of the distance between observed and model-predicted node marginal probabilities \cite{wilinski2021prediction}.
This choice is motivated by the fact that marginal probabilities of nodes' activation can be estimated in a computationally efficient way using a method known as dynamic message passing (DMP) \cite{neri2009cavity, karrer2010message, kanoria2011majority, aurell2011three, aurell2012dynamic, altarelli2013large, altarelli2013optimizing, lokhov2014inferring, lokhov2015dynamic, del2015dynamic, shrestha2015message, lokhov2017optimal, lokhov2019scalable, barthel2020matrix, sun2021competition, li2021impact, aurell2023closure, crotti2023matrix, PhysRevX.13.031021, ghio2023bayes}.
DMP is an inference method for spreading processes on networks derived from a classical belief propagation (BP) algorithm \cite{yedidia2003understanding, mezard2009information}, and thus showing similar properties to BP \cite{lokhov2015dynamic, lokhov2019scalable}. The key result of \cite{wilinski2021prediction} stated that the model learned using the minimization of distance between marginal probabilities generates better predictions compared to the model with ground-truth parameters when DMP is used as the inference algorithm. This has been shown through an empirical study on a popular spreading model known as Independent Cascade \cite{goldenberg2001talk, kempe2003maximizing}, equivalent to a popular epidemic spreading susceptible-infected-recovered (SIR) model \cite{kermack1927contribution, keeling2005networks} with deterministic recovery. Due to a prediction-centric focus, the algorithm proposed in \cite{wilinski2021prediction} has been referred to as SLICER (Scalable Learning of Independent Cascade Effective Representation).

On many sparse network instances, SLICER produces high-quality estimates of model parameters in a time scaling linearly with the system size, even for fraction of hidden nodes up to 25\% of the network. However, from the results of \cite{wilinski2021prediction}, it is not immediately obvious what the expected accuracy limits of the method are, i.e., what error can be expected when the number of unobserved nodes reaches very high values. In order to get insight into the performance of SLICER, in section \ref{sec:learning_limits} we run a systematic evaluation of the performance of SLICER for learning of the spreading model parameters on a variety of random graph classes under diverse fractions of unobserved nodes.

Partial node-observability of the system is only one of the several ways in which the data may be incomplete or uncertain. Some of the challenges often encountered in real data include model selection under the scenarios of unknown or partially known network structure, noisy data, missing observations in time, or necessity to include prior information in order to minimize the required amount of data. Scenarios that we consider are presented in Figure~\ref{fig:scenarios}. The first challenge that we address is the lack of information about the network structure. This task has been recently studied in \cite{huang2022reconstructing}, however, the proposed method involves cubic complexity in the number of nodes, making its application to large networks prohibitive. The second challenge deals with incorporation of prior information on the model parameters in the case where the number of available samples is small.
For instance, in epidemiological applications, at most a few realizations of the dynamics can be observed, but the spreading is often modeled by a single parameter (transmission probability) \cite{daley2001epidemic}.
We show how to leverage this prior information and learn the simple graph models from a few observed trajectories. The third challenge is missing data in time, where nodes report their activation times only during limited time windows. This task was analysed in \cite{sefer2015convex}, but with an additional assumption that the full probabilistic trace for each node is available. A likelihood-free approach for the missing time case was proposed in \cite{dutta2018bayesian}, together with additional goal of finding the source of spreading. A variant of the scenario of missing timestamps assumes that only the complete final state of the dynamics is available. One of the first results for this variant of the problem assumed that the unknown network is a tree \cite{amin2014learning}, but later this assumption has been relaxed in \cite{supeesun2017learning, braunstein2019network, han2020statistical}. In our work, we show that the spreading model can be accuractly recovered with minimal available information in the temporal space, even in the case of partial node-observability. The last challenge that we consider, often related to the reporting procedure, is uncertainty in timestamps, which may be modeled as an additional noise added to the observed data. Similar setting was analysed in \cite{hoffmann2019learning, trouleau2019learning}, but not paired with any other type of missing information such as partial node-wise reporting.

As we show in this work, in most of these scenarios, the objective function used in SLICER method is not directly applicable, resulting in biased estimates of the model parameters. Our overreaching goal consists in generalizing SLICER, introducing a universal learning framework capable of addressing all these challenges. Here, we show how to crucially define the objective function in such a way that it incorporates the available prior information, and provides a high-quality reconstruction of the model parameters. As demonstrated below and contrary to the solution of these challenges on a case-by-case basis, this universality presents an advantage in that data uncertainties present in the data and pertaining to different challenges can be treated at the same time. As a demonstration of such a universality, in the end of our study, we illustrate the approach on real-world network instances under a combination of these challenges.

\section{Methods}

In the section, we first pose the learning problem using the Independent Cascade (IC) model as the dynamics of choice. Then, we discuss the dynamic message passing (DMP) inference method that lies at the foundation of our approach. Finally, we explain the details of our learning method.

\subsection{Problem formulation: Learning of the Independent Cascade model}

In this paper, we focus on the Independent Cascade model \cite{goldenberg2001talk, kempe2003maximizing}.
Define the spreading network as a graph $G = (V, E)$, consisting of the set of vertices $V=\{v_i\}$, and the set of edges $E=\{(v_i, v_j) \, | \, v_i,v_j \in V\}$.
Each node $v_i$ in the IC model can be in one of two states: active and inactive.
Assuming node-wise independent initial condition for the states of the nodes at time $t=0$, which can be either deterministic or stochastic, further dynamics is subject to a single stochastic rule.
If node $i$ gets activated at any time $t$, it has only one chance to activate any of its inactive neighbors $j$ at time $t+1$.
This happens independently for each of the inactive neighbor with edge dependent probability $\alpha_{ij}$:
\begin{equation}
    A(i) + I(j) \xrightarrow{\alpha_{ij}} A(i) + A(j).
\end{equation}
Regardless of activating any of its neighbors, node $i$ remains active forever, but it cannot activate any other node in the future beyond $t+1$.
In what follows, we assume that $G$ is undirected and there are no self-loops or multi-links (although this assumption can be relaxed in general).
The dynamics continues until a predefined number of steps $T$.
A single realisation of this process is called a cascade.
Since each node can be activated only once, every cascade $c$ is fully described by a set of activation times $\{\tau^c_i\}_{i \in V}$.
Additionally, if node $i$ does not get activated before or at the specified time $T$ in cascade $c$, we assign $\tau^c_i=*$ as its activation time.
It means that $*$ summarizes all future activity of a given node.
More details about this notation is given in the Appendix \ref{app:bp}.
In all our experiments, we use a single seed as an initial condition (this setting corresponds to the most popular scenario, but can be relaxed as well).
This means that there is only one node activated at the beginning of the spreading process, although it may be a different node for each specific cascade.
The number of observed cascades is denoted by $M$.

The problem that we consider in this work is as follows: given a set of observed activation times
$\{\tau^c_i\}_{i \in O, c \in C}$ where $O \subset V$ is the set of observed nodes and $C$ is the set of cascades, as well as using additional information on the cascades (e.g., prior knowledge that $\tau^c_i$ are noisy), learn model parameters, i.e., estimate $\{\alpha_{ij}\}_{(ij) \in V\times V}$ so that they are close to the parameters of the ground-truth model, denoted as $\{\alpha^{*}_{ij}\}_{(ij) \in E}$. In the case of the unknown network, thresholded values of $\{\alpha_{ij}\}_{(ij) \in V\times V}$ away from small values close to zero define the recovered network structure $\widehat{E}$ that can be compared to the ground-truth set of edges $E$. In what follows, the fraction of hidden nodes $\vert V \backslash O \vert / \vert O \vert$ is denoted as $\xi$.

We chose the IC model for simplicity reasons.
On one hand, it does capture basic properties of many known spreading processes, including a possibility of an early cascade termination.
On the other, it simplifies the analytical equations presented in the next subsection, making the proposed approach easier to follow and understand.
Our approach can be can easily be generalized to a broad class of more complex spreading models on networks for which DMP equations (explained next) are known.

\subsection{Inference method: Dynamic Message Passing}
\label{sec:dmp_equations}

As discussed in the Introduction, presence of partial information warrants learning methods that feature inference algorithms for predicting the model dynamics as a subroutine. One of the key observables that quantify the spread is given by the \emph{influence function}: the number of expected number of activated nodes at a certain time $t$ for a given initial condition. For a fixed initial conditions in cascade $c$, the influence function is given by the sum of marginal probabilities $p^c_i(t)$ of activation of each node $i \in V$ at at time $t$ \cite{lokhov2017optimal}. Prediction of influence function or marginal probabilities is known to be \#P-hard \cite{chen2010scalable, shapiro2012finding}, and hence approximate methods need to be used. A classical way of estimating the influence function consists in using Monte-Carlo simulations.
However, this approach typically requires a large sampling factor to provide a reliable estimate \cite{kempe2003maximizing, chen2009efficient, du2013scalable, cohen2014sketch, lucier2015influence, nguyen2017outward}.
Aiming at an accelerated approach that saves this sampling factor, we use the inference method known as Dynamic Message Passing \cite{neri2009cavity, karrer2010message, kanoria2011majority, aurell2011three, aurell2012dynamic, altarelli2013large, altarelli2013optimizing, lokhov2014inferring, lokhov2015dynamic, del2015dynamic, shrestha2015message, lokhov2017optimal, lokhov2019scalable, barthel2020matrix, sun2021competition, li2021impact, aurell2023closure, crotti2023matrix, PhysRevX.13.031021, ghio2023bayes}.
For IC model \cite{lokhov2019scalable}, DMP estimates marginal probabilities of activation in a linear time in both system size and duration of the dynamics, and has the properties of being exact on graphs without loops and asymptotically exact on random graphs, providing an upper-bound of the influence function on general networks.    

For the IC model, the equations take the following form:
\begin{equation}
    p^c_i(t) = 1 - \big( 1 - \bar{p}^c_i \big) \prod_{k \in \partial i} \! \Big(1 - \alpha_{ki} \cdot p^c_{k \rightarrow i}(t - 1) \Big),
    \label{eq:marginal}
\end{equation}
\begin{equation}
    p^c_{j \rightarrow i}(t) = 1 - \big( 1 - \bar{p}^c_j \big) \prod_{k \in \partial j \setminus i} \!\! \Big(1 - \alpha_{kj} \cdot p^c_{k \rightarrow j}(t - 1) \Big),
    \label{eq:message}
\end{equation}
where $p^c_i(t)$ is the marginal probability of node $i$ being active at time $t$ under initial conditions of cascade $c$, $p^c_{i \rightarrow j}(t)$ is the same probability, but on an auxiliary graph where node $j$ was removed, $\partial i$ denotes the set of neighbors of node $i$ in the graph $G$, and $\partial j \setminus i$ denotes the set of neighbors of node $j$ in the graph $G$ except $i$.
We also denote an initial condition for node $i$ in cascade $c$ as $\bar{p}^c_i = p^c_i(0)$.
Note that for known model parameters, the marginals given by DMP depend only on the initial condition $\bar{p}^c_i$. Detailed properties of the above equations were studied in \cite{lokhov2019scalable}. In Appendix~\ref{app:bp}, following the approach of \cite{lokhov2015dynamic}, we provide an alternative derivation of DMP equations for IC model that connects them and their properties to the classical belief propagation algorithm \cite{yedidia2003understanding, mezard2009information}.

\subsection{Learning Framework}

In this section, we explain the details of our method. We start by stating the approach introduced in \cite{wilinski2021prediction} and known as Scalable Learning of Independent Cascade Effective Representation (SLICER), which will be used as a baseline method in all numerical experiments below. Subsequently, we will show that the objective used in SLICER need to be modified to account for the prior information on the uncertainty in the data, and enhanced with the known constraints on the model parameters -- an approach referred to as SLICER+.

Let us focus on the case of perfectly observed information from a subset of visible nodes $O$ \cite{wilinski2021prediction}. SLICER is based on the minimization of the Kullback–Leibler (KL) divergence between the empirical marginal distributions computed from the data, and the marginals estimated from the model using DMP, on the observed nodes. Marginal distributions depend on non-local model parameters, including those adjacent to hidden nodes, and thus allowing for the reconstruction of the entirety of model parameters under sufficient observations. Minimization of KL distance on marginals on the observed nodes in our case is equivalent to maximizing:
\begin{equation}
    \mathcal{O} = \sum_{c \in C} \sum_{i \in O} \log \mu^c_i(\tau^c_i),
    \label{eq:cost_function}
\end{equation}
where $C$ is the set of available cascades, $O$ is the set of visible nodes and $\mu^c_i(t)$ is the marginal probability of node $i$ being activated under cascade $c$ precisely at time $t$.
These marginal probabilities can be calculated based on the marginal variables $p^c_i(t)$ used in the DMP equations above:
\begin{equation}
    \mu^c_i(t) = p^c_i(t) \cdot \mathds{1}_{(t < T)} - p^c_i(t - 1) \cdot \mathds{1}_{(t > 0)} + \mathds{1}_{(t = T)},
\end{equation}
where $\mathds{1}$ stands for the indicator function.

The cost function \eqref{eq:cost_function} has been first proposed in \cite{lokhov2016reconstructing} showing an asymptotic consistency of DMP-based recovery using this objective, but providing an inefficient optimization algorithm. The work \cite{wilinski2021prediction} proposed an efficient algorithm for minimizing \eqref{eq:cost_function}, which gave rise to the SLICER algorithm.

For a given set of parameters $\alpha_{ij}$, marginal probabilities depend only on the initial condition.
Therefore, we can re-write the objective in the following way:
\begin{equation}
    \mathcal{O} = \sum_{s \in S} \sum_{i \in O} \sum_{\tau_i^s} m^{\tau^s_i} \log \mu^s_i(\tau^s_i),
    \label{eq:obj}
\end{equation}
where $S$ is a set of all the initial conditions occurring across all the cascades $C$ and $m^{\tau^s_i}$ is the number of times node $i$ gets activated at time $\tau^s_i$ under initial condition $s$.
This equivalent reformulation allows one to reduce the computing cost by evaluating DMP equations only $|S|$ times instead of $M=|C|$.
In our simulations, where $s$ is assumed to be a single node, it means that DMP will not be run more than $N$ times, regardless of the number of available cascades.
In the case of a stochastic initial condition of all of the cascades, e.g., when each node is independently initially activated with probability $\frac{1}{N}$, the size of $S$ is $|S|=1$, making the whole computation far less costly.

In order to maximize (\ref{eq:obj}) we use a Lagrangian formulation, where the objective function $\mathcal{O}$ \eqref{eq:obj} is supplemented with constrains on marginal provided via DMP,
\begin{equation}
    \mathcal{L} = \!\!\underbrace{\mathcal{O}}_{objective} + \!\!\!\underbrace{\mathcal{C}}_{constraints}\!\!\!\!\!,
    \label{eq:Lagrangian}
\end{equation}
where in the absence of any additional prior information on the parameters, $\mathcal{C}$ are given by DMP equations (\ref{eq:marginal})-(\ref{eq:message}) reweighted by Lagrange multipliers $\lambda^s_i(t)$ for all nodes $i \in V$ and $\lambda^s_{i \rightarrow j}(t)$ for all edges $(ij) \in E$ of graph $G$ for each time $t$ and for each cascade in the class $s$, see Appendix~\ref{app:SLICER+_derivation} for details.
The iterative equations constituting SLICER is obtained by differentiating the Lagrangian \eqref{eq:Lagrangian} with respect to all variables: ${p_i^s(t)}_{i \in V}$, ${p_{i \to j}^s(t)}_{(ij) \in V \times V}$, ${\alpha_{ij}}_{(ij) \in V \times V}$, ${\lambda_i^s(t)}_{i \in V}$, and ${\lambda_{i \to j}^s(t)}_{(ij) \in V \times V}$. This results in DMP equations (\ref{eq:marginal})-(\ref{eq:message}) in the primal space; DMP-like equations on the Lagrange multipliers ${\lambda_i^s(t)}_{i \in V}$, and ${\lambda_{i \to j}^s(t)}_{(ij) \in V \times V}$ in the dual space; and update equations for the parameters ${\alpha_{ij}}_{(ij) \in V \times V}$, see Appendix~\ref{app:SLICER+_derivation}.

When prior information about the parameters is available, it can be directly incorporated in the term $\mathcal{C}$. As an example, the constraint can be greatly simplified if we know that for each $(ij) \in E$, $\alpha_{ij} = \alpha$ (the case of so-called simple graphs, representing a popular case in epidemiological models). As we show below in the section \ref{sec:simple_graphs}, the gradient of the Lagrangian with respect to the model parameter $\alpha$ in this case reads for $\alpha \neq 0$:
\begin{equation}
    \frac{\partial \mathcal{L}}{\partial \alpha} = -\frac{1}{\alpha} \sum_{s \in S} \sum_{t=0}^{T-1} \sum_{(i, j) \in V \times V} \lambda^s_{i \rightarrow j}(t) \cdot p^s_{i \rightarrow j}(t).
    \label{eq:alpha}
\end{equation}
The gradient in Eq.~\eqref{eq:alpha} can be used to learn parameter $\alpha$ using an iterative procedure with $\alpha \longleftarrow \alpha + \varepsilon \frac{\partial \mathcal{L}}{\partial \alpha}$,
where $\varepsilon$ is a learning rate.
Lagrangian formulation ensures that a single gradient descent step has the worst-case computational complexity $O(|E| T |S|)$, which means it is linear in the system size, cascade length and the number of initial conditions.

In a similar manner, under different scenarios of incomplete data considered in Figure~\ref{fig:scenarios}, the loss function $\mathcal{O}$ can be appropriately modified to include the prior information $A$ on the type of the uncertainty in the data, as we discuss in the next section below. Collectively, we refer to the algorithm based on modified objective $\mathcal{O}$ and constraints $\mathcal{C}$ as to SLICER+. The full derivation of SLICER+ for each of the challenges in Figure~\ref{fig:scenarios} is given in the Appendix \ref{app:SLICER+_derivation}. Efficient implementation of the algorithm, taking advantage of the DMP-like equations on the evolution of the Lagrange multiplies in the dual space and assuring the linear complexity of the algorithm, is available in the Appendix \ref{app:eff}.

\begin{figure*}[!htb]
    \includegraphics{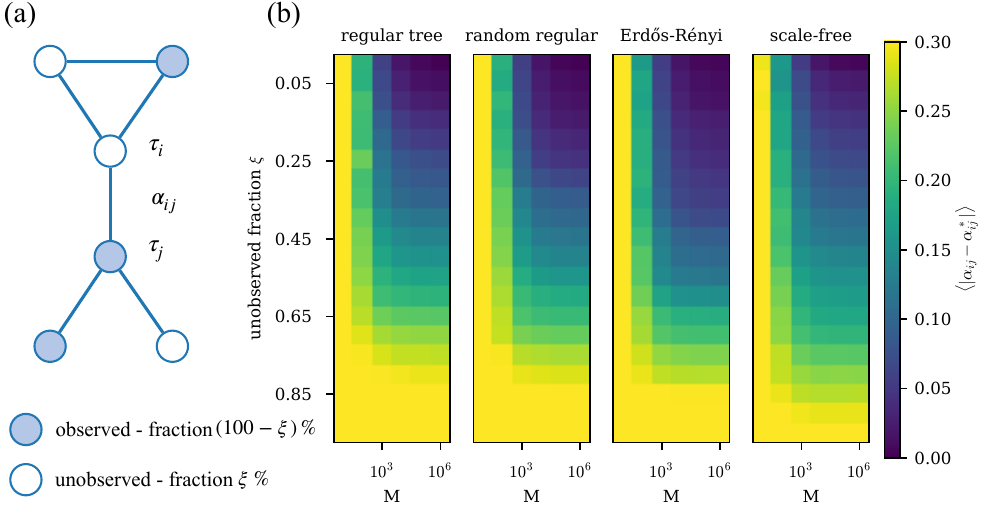}
    \caption{\label{fig:heatmap} Heat maps of average difference between inferred $\alpha_{ij}$ and real $\alpha^*_{ij}$ parameters in $\ell_1$ norm, as a function of the number of available cascades and the fraction of unobserved nodes.
    Each heat map represents a different network type, but all of them have the same color scale.
    Each point is averaged over 5 different networks and 5 different sets of parameters $\alpha^*_{ij}$ (sampled from a uniform distribution in the range $[0,1]$).
    All networks contain $N=100$ nodes and all but the tree have average degree $\langle k \rangle = 3$.
    Unobserved nodes were picked at random.
    All cascades had length $T=5$.
    Note that $\xi = 33\%$ corresponds to a random guess.}
\end{figure*}

\section{Results}

In this section we present details on the SLICER+ approach designed for dealing with different types of challenges regarding uncertainty and partial observability of data presented in Figure~\ref{fig:scenarios}.
All subsections devoted to particular challenges are supported with simulations on synthetic networks.
For this purpose we choose four different network models: 3-regular tree (RT), 3-regular random graph (RR), Erd\"{o}s-R\'{e}nyi graph with average degree equal to 3 (ER) \cite{bollobas1998random} and a scale-free graph generated with Barab\'{a}si-Albert model with average degree equal to 3 (BA) \cite{barabasi1999emergence}.
Except for the tree, all models have the same number of edges, so that they are comparable.
The tree network case is studied for the baseline benchmarking purposes given that DMP is exact on trees.
In the section~\ref{sec:real-world}, we apply the method to real-world networks, while assuming a simultaneous co-occurence of multiple types of incomplete data.

\subsection{SLICER's learning limits}
\label{sec:learning_limits}

SLICER was already shown in \cite{wilinski2021prediction} to perform well in a regime where the unobserved part of the system is significant.
It was not shown, however, what are the limits of the algorithm in terms of the size of the unobserved part and how it depends on the sample size.
Here we answer this question and additionally we show how these limits depend on the network structure.

\begin{figure*}[!htb]
    \includegraphics{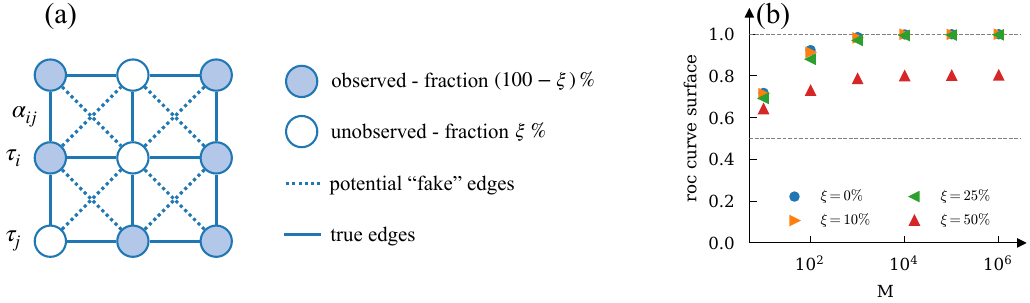}
    \caption{\label{fig:lattice}
    (a) The structure of the lattice expanded by additional fake (inactive) edges.
    The structure learning task is to find the true edges, which were used to produce the observed cascades.
    (b) Average ROC curve surface, as a function of the number of available cascades for square lattice with additional fake (inactive) edges in the case where a fraction $\xi$ of nodes is unobserved.
    Each point is averaged over five different sets of parameters $\alpha^*_{ij}$ (sampled from a uniform distribution in the range $[0,1]$).
    Network contains $N=100$ nodes.
    Unobserved nodes were picked at random.
    All cascades had length $T=5$.}
\end{figure*}

We take all four synthetic network models, generate up until $M=10^6$ cascades and vary the percentage of the unobserved part from 0\% to 95\% (with a 5\% step).
The results of applying SLICER to all these cases are presented in the form of heat maps in Fig. \ref{fig:heatmap}.
Interestingly, the quality of reconstructed parameters decrease faster for tree graphs than for loopy networks, because tree graphs have significantly lower edge density than other networks, making it not directly comparable.
At the same time, single unobserved node affects a tree structure in a more significant way -- single node separates a tree into two disconnected sub-graphs.
We observe an interesting behavior from a scale-free graph.
Despite a steeper decline in the quality of reconstruction for smaller fractions of the unobserved part $\xi$, for high values of $\xi$ the reconstruction still correlate with the true solution.
If we look at $80-90\%$ of unobserved nodes for both random regular and Erd\"{o}s-R\'{e}nyi graph, the reconstructed parameters are basically random.
Dynamics on scale free graph on the other hand, still does contain information that one is able to exploit, even for unobserved fraction equal to $\xi=90\%$.
For lower values of the unobserved fraction $\xi$, in the interval between $30-70\%$, the best parameters' reconstruction results are obtained for the  Erd\"{o}s-R\'{e}nyi graph.
Finally, when $\xi$ is below $30\%$, the best parameters' reconstruction is achieved for the regular random structure.
These results illustrate that the uncertainty of the obtained reconstruction depends on the particular network structure.

\begin{figure*}[!htb]
    \includegraphics{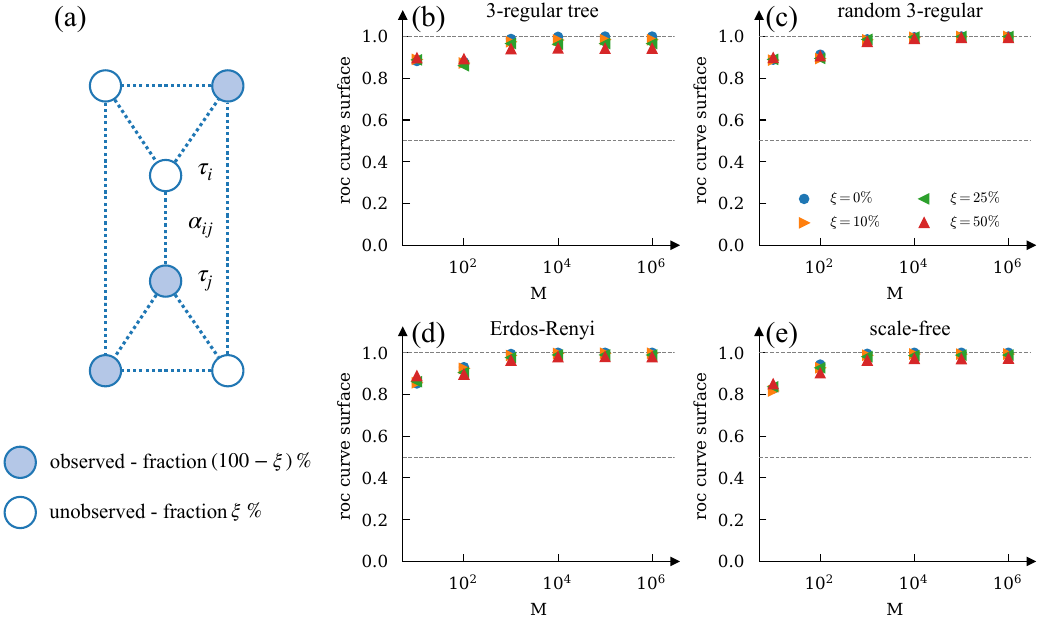}
    \caption{\label{fig:struct} Structure learning starting with a super-set of edges. Average ROC curve surface, as a function of the number of available cascades for different network types in the case where a fraction $\xi$ of nodes is unobserved and there is a known set of potential edges, where the number of \textit{fake} edges is equal to the \textit{true} ones.
    Each point is averaged over five different networks and five different sets of parameters $\alpha^*_{ij}$ (sampled from a uniform distribution in the range $[0,1]$).
    All networks contain $N=100$ nodes and all but the tree have average degree $\langle k \rangle = 3$.
    Unobserved nodes were picked at random.
    All cascades had length $T=5$.}
\end{figure*}

\subsection{Structure learning}

In many real-world applications, like epidemic spreading, data on contact network is extremely difficult to gather.
Sometimes it is possible to assess a limited number of potential connections, but detailed knowledge about the graph is rarely available.
In practice, structure learning is equivalent to learning spreading parameters, where near-zero couplings signal an absence of an edge. Under the structure recovery task, parameter learning needs to be run on larger graph reflecting the prior knowledge on the network structure.
The less knowledge one have about the network, the larger the set of edges one needs to consider, up until $N(N-1)/2$ potential edges in a fully-connected graph when no knowledge about model structure is provided.
When the dynamics of all nodes is available, edges can be recovered using maximum likelihood approach and thresholding.
We focus on a more challenging scenario where the network structure is unknown or partially known under the presence of unobserved nodes.
One of the main difficulty of this scenario consists in an increased potential for the solution degeneracy.
Indeed, already in the case of full structural knowledge, degeneracy may appear for specific arrangements of unobserved nodes.
The simplest example is when an unobserved node is a leaf -- a node with a single connection.
In this case it is impossible to recover the outgoing spreading parameter corresponding to the leaf edge.
There are also more complex situations involving solution degeneracy such as interconnected clusters of hidden nodes, and increased number of unknown connections makes these situations more likely.
However, information contained in the cascades makes it possible to discover the structure of the diffusion network with high accuracy even on very loopy graphs, as we show below.

Consider a problem of selecting the spreading graph from a set of known super-set of edges. Such a super-set of edges becomes a fully-connected graph in the worst case of no prior information on the diffusion network. SLICER can then be used as the structure discovery algorithm as follows. A gradient descent on the parameters $\{ \alpha_{ij} \}$ as a part of SLICER is run on a super-set of edges. If a parameter goes beyond a certain threshold value (which we take as $10^{-8}$ in all experiments in this section), it is declared as zero and the respective edge is removed from the set of candidate edges, thus reducing the network at the next steps of the gradient descent procedure. This online pruning procedure results in a reduced overall computational complexity of the algorithm, compared to an alternative where SLICER is run on the full super-set of edges and the parameters are thresholded only in the end of the procedure. In order to evaluate the effectiveness of our method we use a receiver operating characteristic (ROC) curve \cite{fawcett2006introduction}. After assessing all potential edges, we build the ROC curve of true and false positives among them and then we compute the surface under such curve. ROC curve surface equal to $1$ represents a perfect reconstruction, while values oscillating around $0.5$ suggest that the edge set selection is not better than a random guess.

Surprisingly, structure discovery with SLICER using the procedure described above is successful even when the underlying network has many short loops, which strongly affects the accuracy of the DMP approximation, as explained in section~\ref{sec:dmp_equations}. To illustrate this point, consider an adversarial scenario of a regular two-dimensional square lattice containing a large number of short loops, which represents the ground-truth propagation network. Further, consider a super-set of edges by adding \textit{diagonal} connections to the lattice, as shown in Fig. \ref{fig:lattice}(a).
Thus expanded network containing $50\%$ fake edges (and even more short loops compared to the ground-truth square lattice) is then used as a starting network in the structure learning task.
Ground-truth parameters $\alpha_{ij}$ are drawn from a uniform distribution over $[0, 1]$ interval. Despite this adversarial scenario, we are able to perfectly reconstruct the correct network, see Fig.~\ref{fig:lattice}, even with $25\%$ of nodes being unobserved.

In Fig.~\ref{fig:struct}, we test structure recovery procedure on four different types of synthetic networks, where the size of the super-set of edges is two times larger than the number of ground-truth edges in the network on which the data has been generated. These results show that an accurate recovery of the network structure is possible even in the presence of a large fraction of hidden nodes. In Section~\ref{sec:real-world}, we further explore structure learning for real-world network instances. In Appendix~\ref{app:topo}, we discuss results for network structure recovery in the case of no available prior information on the topology, where the super-set of edges corresponds to a fully-connected graph.

\begin{figure*}[!htb]
    \includegraphics{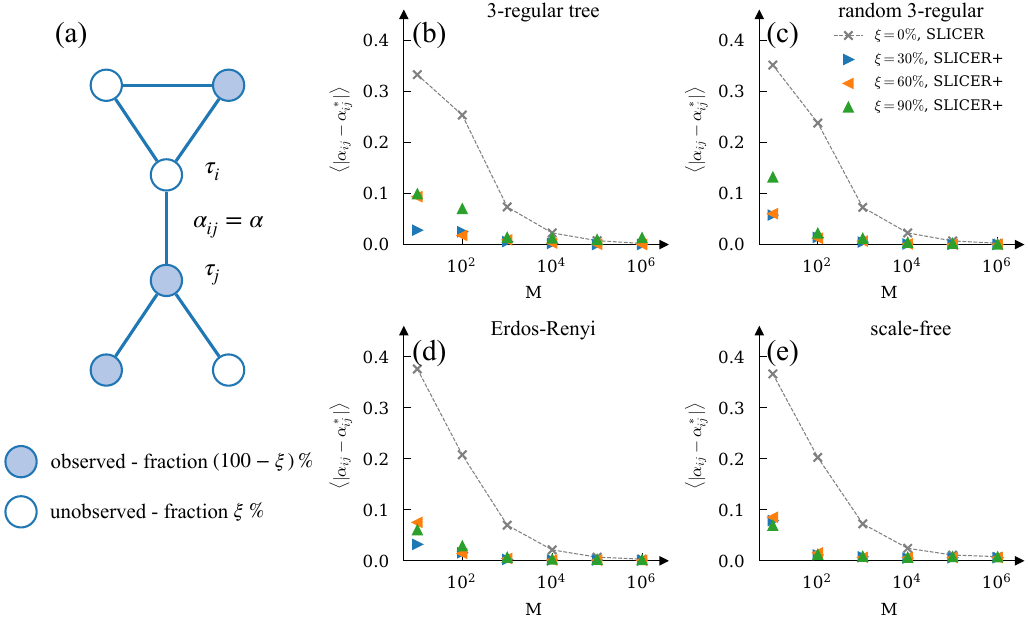}
    \caption{\label{fig:simple} Learning of simple graphs with equal transmission probabilities on all edges. Average difference between inferred $\alpha_{ij}$ and real $\alpha^*_{ij}$ parameters in $\ell_1$ norm, as a function of the number of available cascades for different network types in the case where a fraction $\xi$ of nodes is unobserved.
    The dashed gray line denotes the benchmark case where one is using standard SLICER without imposing the constraint of equal parameters on all edges.
    The triangles corresponds to the case where SLICER includes the knowledge about equal parameters.
    Each point is averaged over 5 different networks with parameters $\forall_{(i, j) \in E} \, \alpha^*_{ij} = 0.5$.
    All networks contain $N=100$ nodes and all but the tree have average degree equal to $\langle k \rangle = 3$.
    Unobserved nodes were picked at random.
    All cascades had length equal to $T=5$.}
\end{figure*}

\subsection{Simple graphs}
\label{sec:simple_graphs}

In many applications, some knowledge about model parameters is assumed, which makes it easier to estimate them.
Most often it is simply assumed that all the spreading parameters are the same for all edges $\alpha_{ij} = \alpha \, \forall_{(i,j) \in E}$.
Although such setting significantly simplifies the problem, it is a good case study for understanding how the algorithm can take advantage of additional knowledge.

The constraint of equal parameters change the DMP equations, which leads to a different form of the SLICER+ instantiation for this case.
In this new setting the constraints become a single parameter function:
\begin{equation}
    \mathcal{L} = \underbrace{\mathcal{O}(\{\tau^c_i\})}_{objective} + \underbrace{\mathcal{C}(\alpha)}_{constrains},
\end{equation}
which requires recomputing the Lagrangian derivatives.
Following the steps described in the Appendix \ref{app:simple_form} we arrive at:
\begin{equation}
    \frac{\partial \mathcal{L}}{\partial \alpha} = -\frac{1}{\alpha} \sum_{s \in S} \sum_{t=0}^{T-1} \sum_{(i, j) \in E'} \lambda^s_{i \rightarrow j}(t) \cdot p^s_{i \rightarrow j}(t).
\end{equation}
In the end, not only do we reduce the memory usage of the algorithm, but most importantly, we reduce the amount of data needed to obtain desirable level of error on parameters.

We apply the modified procedure to synthetic data generated with different network types.
As shown in Fig. \ref{fig:simple} even with $90\%$ unobserved nodes and regardless of the network structure, a small number of cascades is needed to obtain remarkable accuracy.
Ten cascades are on average enough to get error below $0.1$ and it does not depend strongly on the size of the unobserved part.
Assuming different transmission probabilities for each edge in the reconstruction, shown with a gray dashed line as a benchmark, yields results which are orders of magnitude worse.
Moreover, as shown in Fig. \ref{fig:heatmap}, different transmission probabilities combined with unobserved fraction above $60\%$ yields almost random outcome, even with large number of cascades.

\begin{figure*}[!htb]
    \includegraphics{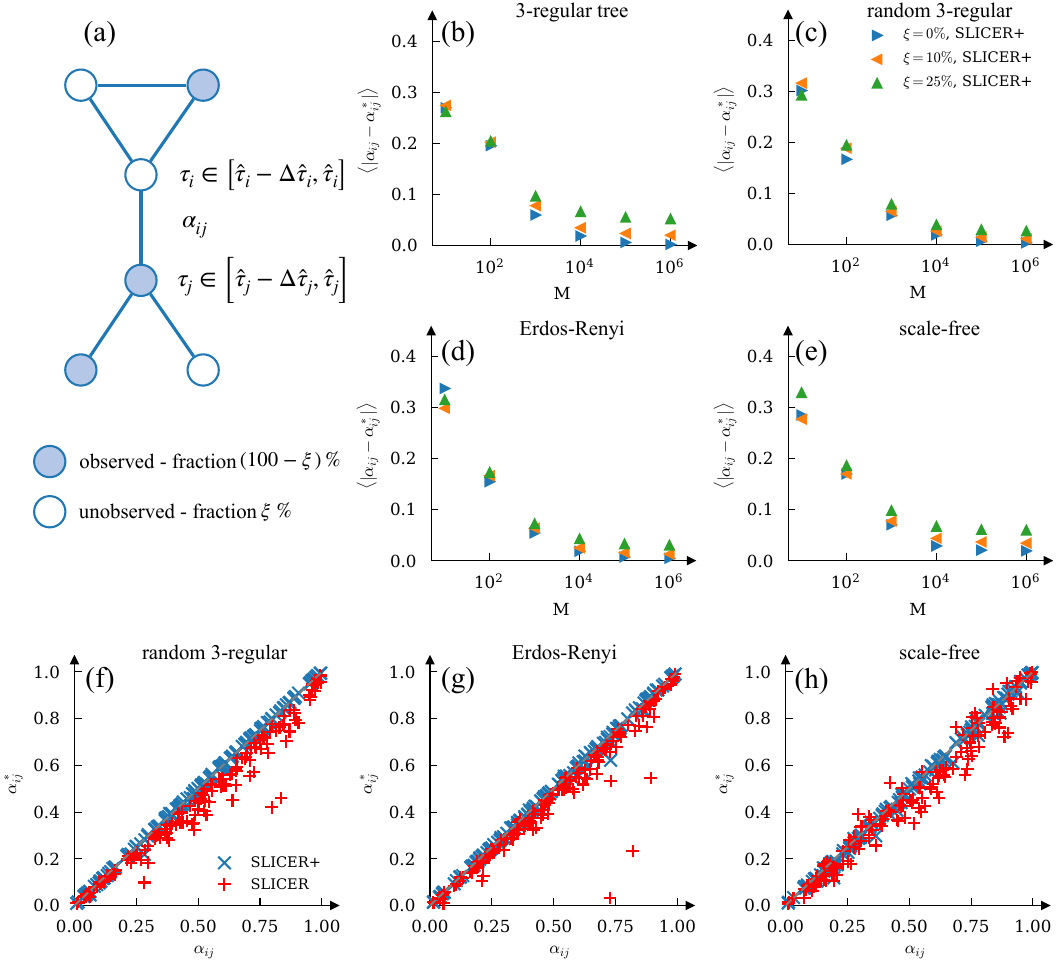}
    \caption{\label{fig:unobs_time}
    \textbf{(a)} Reconstruction of parameters under missing temporal observations. \textbf{(b-e)} Average difference between inferred $\alpha_{ij}$ and real $\alpha^*_{ij}$ parameters in $\ell_1$ norm, as a function of the number of available cascades for different network types.
    Only initial and final spreading states are observed and additionally, some nodes are not observed at all.
    The triangles corresponds to the case where SLICER includes the knowledge about unobserved periods of time.
    Each point is averaged over 5 different networks and 5 different sets of parameters $\alpha^*_{ij}$ (sampled from a uniform distribution in $[0,1]$).
    All networks contain $N=100$ nodes and all but the tree have average degree equal to $\langle k \rangle = 3$.
    Unobserved nodes were picked at random.
    All cascades had length equal to $T=6$.
    \textbf{(f-h)} Scatter plots of true parameters $\alpha^*_{ij}$ versus the estimated ones $\alpha_{ij}$.
    The results were obtained for different network structures with $N=100$ nodes, $M=10^6$ cascades of length $T=6$ and $\xi^{time} = 33.(3)\%$ of unobserved times picked at random.
    Except for the tree, all networks have average degree equal to $\langle k \rangle = 3$.}
\end{figure*}

\subsection{Missing temporal observations}

Not being able to observe the whole system, as a result of a limited budget or other challenges related to gathering data, is a fair limitation for many practical applications.
Another realistic situation arises when, for similar reasons, one is not able to record the data for extended periods of time.
Partial observability of the system, should be considered in both spatial and temporal spaces.
Here we show how to modify the objective so that it correctly accounts for the unobserved time periods, resulting in a SLICER+ instantiation for the scenario of missing observations in times.

Under this scenario, the state of the network is not observed at each time step, but only at a subset of observation times. Therefore, although we do not generally observe activation time of a node, we still know that activation happened somewhere in the last unobserved period preceding the first time we noticed a given node to be active.
As a result there are two cases: (i) node activation is observed and we denote the activation time as $\tau_i$; (ii) node activation happens during an unobserved interval denoted as $[\hat{\tau}_i - \Delta \hat{\tau}_i, \hat{\tau}_i]$. Observed snapshots of the initial and of the final states of the spreading dynamics only represents a particular example of such a missing information in time. 
New objective will depend on known activation times and activation intervals:
\begin{equation}
    \mathcal{L} = \underbrace{\mathcal{O}(\{\tau^c_i\}, \{\hat{\tau}_i, \Delta \hat{\tau}_i\})}_{objective} + \underbrace{\mathcal{C}(\{\alpha_{ij}\})}_{constrains}.
\end{equation}
This leads to a variant of the SLICER+ algorithm for missing temporal observations, presented in Appendix \ref{app:times_form}.

\begin{figure*}[!htb]
    \includegraphics{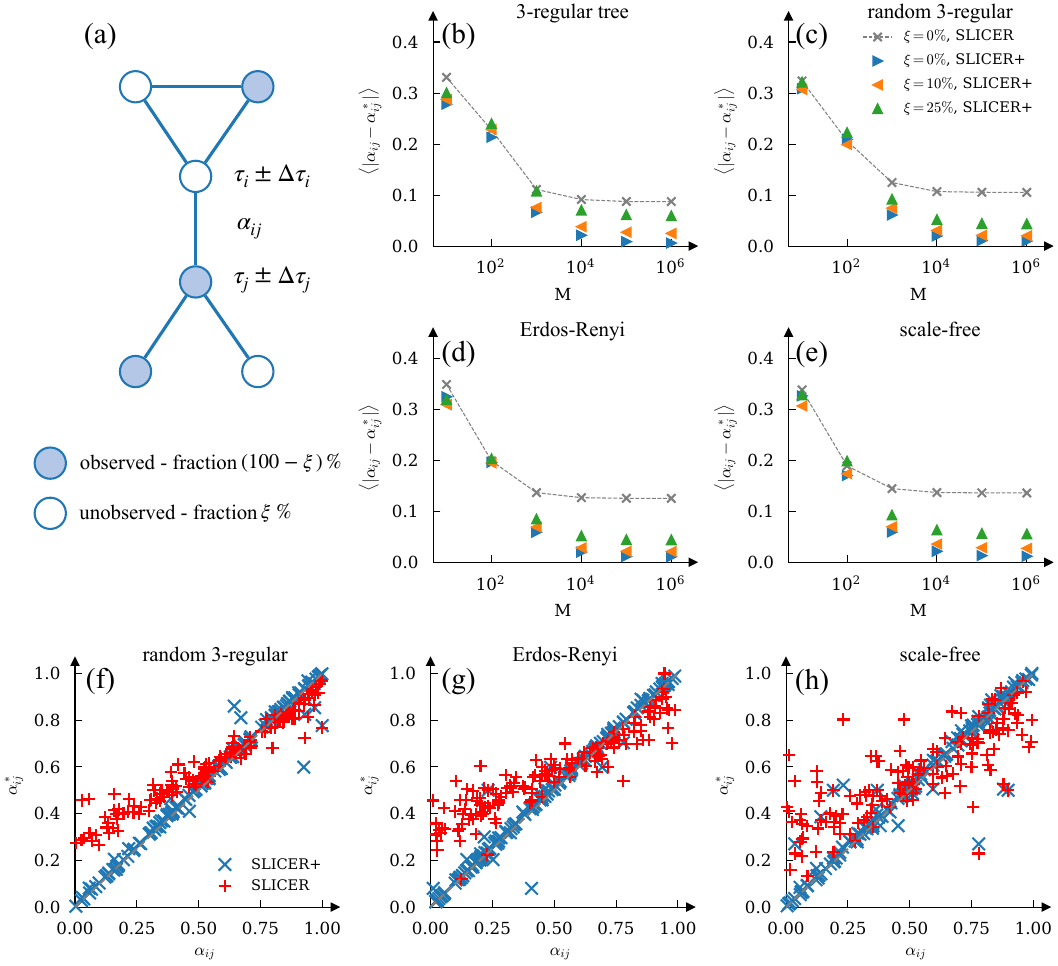}
    \caption{\label{fig:noisy}
    \textbf{(a)} Reconstruction of parameters under noisy observations. \textbf{(b-e)} Average difference between inferred $\alpha_{ij}$ and real $\alpha^*_{ij}$ parameters in $\ell_1$ norm, as a function of the number of available cascades for different network types in the case where a fraction $\xi$ of nodes is unobserved.
    Additionally, observed activation times $\{ \tau_i^c \}_{i \in O}$ are subject to noise described by Eq. \ref{eq:synth_noise}.
    The dashed gray line denotes the benchmark case obtained with SLICER.
    The triangles corresponds to the case where SLICER includes the knowledge about the noise.
    Each point is averaged over 5 different networks and 5 different sets of parameters $\alpha^*_{ij}$ (sampled from a uniform distribution in $[0,1]$).
    All networks contain $N=100$ nodes and all but the tree have average degree equal to $\langle k \rangle = 3$.
    Unobserved nodes were picked at random.
    All cascades had length equal to $T=5$.
    \textbf{(f-h)} Scatter plots of true parameters $\alpha^*_{ij}$ versus the estimated ones $\alpha_{ij}$.
    The results were obtained for different network structures with $N=100$ nodes, $M=10^6$ cascades of length $T=5$ and $\xi = 10\%$ of unobserved nodes.
    Additionally, observed activation times $\{ \tau_i^c \}_{i \in O}$ are subject to noise described by Eq. \ref{eq:synth_noise}.
    Except for the tree, all networks have average degree equal to $\langle k \rangle = 3$.}
\end{figure*}

Results in Fig. \ref{fig:unobs_time}(b-e) are presented in the case where only initial and final states of the dynamics are observed, similarly to the case considered in \cite{braunstein2019network} (where the case of full spatial observations $\xi = 0$ has been treated).
Given the effectiveness of the method in the setting of unobserved time-periods, we explore an even more complex case, where we combine both unobserved times and unobserved nodes, a realistic setting, which, to our best knowledge, was not addressed in the literature before.
Results presented in Fig. \ref{fig:unobs_time}(b-e) are of similar quality as the ones obtained in \cite{wilinski2021prediction} when there are no missing observations in time, despite lack of knowledge on the exact activation times, which illustrates the robustness of SLICER+ with respect to missing observations in time.

Unobserved time periods can be naively treated with SLICER, by simply assuming that if a given node was activated in such a period, it is regarded as an unobserved node.
This, however, does not take advantage of all the available information.
As it is shown in Fig. \ref{fig:unobs_time}(f-h), this leads to a one-sided bias, which quantitatively depends on the network structure and the fraction of unobserved timestamps.
When the unobserved periods are treated correctly, as done in SLICER+, the bias disappears.

\subsection{Noisy timestamps}

Another realistic scenario in data collection relates to the fact that apart from not being able to monitor everything, the collected data is not necessary perfect.
This may be a result of omission, deliberate actions or simply imperfect gathering procedure.
If, for example, a population is tested for a certain virus, the testing dates are most likely not the dates when the infection occurred.
On top of that, the lab results may be delayed, dates may be mistaken and tests can produce both false positives and false negatives.
All of this builds up into an unpredictable noise, which cannot be removed in a deterministic matter.
It has to be accounted for in the modelling process.

In our case of learning spreading parameters, the data is a set of activation times.
In this context, we model noisy observations as deviations around the correct activation times.
We assume that what is observed is a sum of the original activation time and a noise coming from a certain discrete distribution.
In general it is described as follows:
\begin{equation}
    P(\tau = \tau^* + k) = \pi_k \quad \forall \vert k \vert \leq K,
\end{equation}
where $\tau$ is the observed activation time, $\tau^*$ is the real activation time, and $K$ is a constant related to the support of the noise distribution.
The presence of noise in timestamps has no effect on the DMP equations and the constraints, but will affect the objective function, resulting in incorporation of the prior on the noise in the SLICER+ formulation.
The objective function will now depend on both activation times and the noise distribution:
\begin{equation}
    \mathcal{L} = \underbrace{\mathcal{O}(\{\tau^c_i\}, \{\pi_k\})}_{objective} + \underbrace{\mathcal{C}(\{\alpha_{ij}\})}_{constrains}.
\end{equation}
Exact form of the objective for noisy data and detailed derivation of SLICER+ for this case is given in the Appendix \ref{app:noisy_form}.

As in the previous sections, we test the instantiation of SLICER+ for the case of noisy observations with different network structures.
In the numerical experiments we use noise described by the following distribution for $K=1$:
\begin{equation}\label{eq:synth_noise}
    \pi_k = \left\{ \begin{array}{ll}
        0.6, \quad \text{if } k=0,\\
        0.2, \quad \text{if } |k|=1,\\
        0.0, \quad \text{otherwise}.
    \end{array} \right.
\end{equation}
For simplicity, in the numerical experiments in this section, we assume that the noise distribution is known, but the probabilities $\pi_k$ characterizing the noise distribution can be treated as parameters and learned from the data.

Notice that a naive application of SLICER to the case of noisy observations could lead to infinite objective function, since the realization of noisy timestamps could be \textit{impossible}, showing inconsistency with the dynamic rules of the IC model.
This could happen, for instance, if a node is observed to be activated before any of its neighbor is activated.
Application of SLICER is still possible by neglecting these conflicting cases leading to an infinite value of the objective function. We use such an approach as a benchmark to SLICER+ that accounts for the possibility of noisy observations.
Fig. \ref{fig:noisy}(b-e) shows that the use of SLICER quickly leads to a significant error gap, while consideration of noise implemented in SLICER+ brings the algorithm's performance to a level similar to a noiseless situation.
Looking more closely at the particular results at Fig. \ref{fig:noisy}(f-h), we see that disregarding the presence of noise gives results that are correlated with the true solution, but there is a systematic bias, which depends on the network structure.
The solution given by the noise-corrected algorithm, on the other hand, aligns perfectly with the ground-truth.

\subsection{Study on real-world networks}
\label{sec:real-world}

Previously we used different network structures to show how they affect the results of the learning procedure for different scenarios.
Real-world networks typically combine several characteristics of different random models \cite{newman2003structure}.
In order to truly test the effectiveness of our learning framework we use two real-world social networks together with multiple types of noisy and incomplete spreading data.
The first network is the Zachary Karate Club \cite{zachary1977information} with $N=34$ nodes, $|E|=78$ edges and multiple short loops, small enough to serve as an explicit illustration of the quaility of structure learning under partial and noisy observations.
The second network is a Facebook snapshot from \cite{nr-aaai15} and previously analysed in \cite{traud2012social}, with $N=2888$ nodes and $|E|=2981$ edges.

\begin{figure*}
    \includegraphics{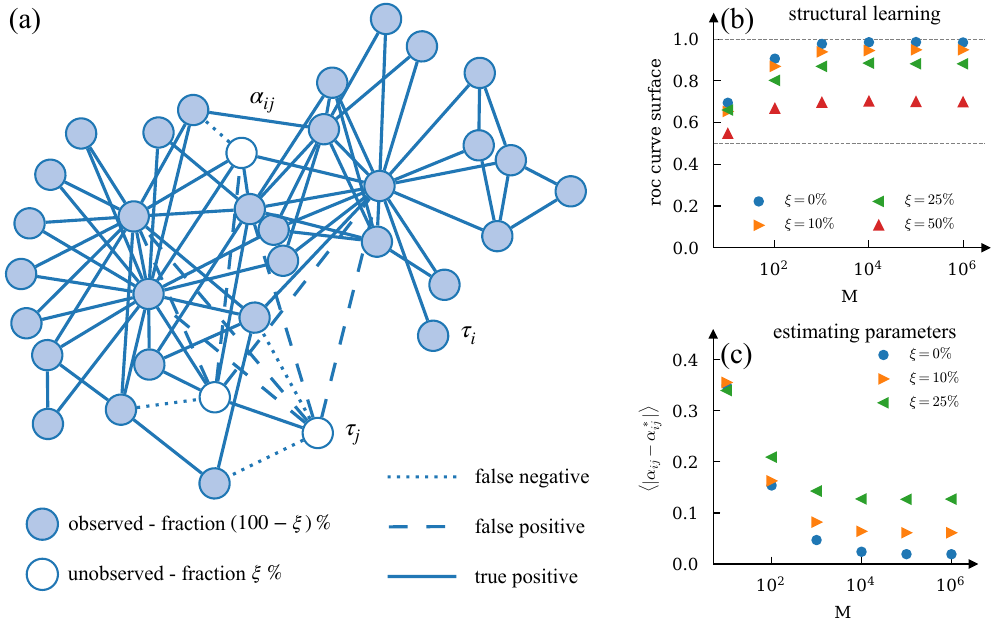}
    \caption{\label{fig:zachary} (a) Structure reconstruction using the Zachary Karate Club network.
    Solid black connections represent the correctly reconstructed edges.
    Dotted black connections are the ground-truth edges that were not reconstructed.
    Dashed gray lines are the ones reconstructed by the algorithm, but absent in the ground-truth edge set.
    We fix all of spreading parameters to $\alpha_{ij} = 0.5$.
    Network was learned based on $M = 1000$ cascades of length $T = 5$ with $\xi = 10\%$ of unobserved nodes.
    (b) Average ROC curve surface, as a function of the number of available cascades for the Zachary Karate Club network in the case where a fraction $\xi$ of nodes is unobserved.
    (c) Average difference between inferred $\alpha_{ij}$ and real $\alpha^*_{ij}$ parameters in $\ell_1$ norm, as a function of the number of available cascades for the Zachary Karate Club network in the case where a fraction $\xi$ of nodes is unobserved.
    In both cases (b) and (c), each point is averaged over 5 different sets of parameters $\alpha^*_{ij}$ (sampled from a uniform distribution in $[0,1]$).
    Network contains $N=34$ nodes and $|E|=78$ edges.
    Unobserved nodes were picked at random.
    All cascades had length $T=5$.
    We assume no knowledge on the topology, hence the initial set of edges is the complete graph.
    Additionally, observed activation times are subject to known noise described in Eq. \ref{eq:synth_noise}.
    }
\end{figure*}

In Figure~\ref{fig:zachary}, we present results on learning of the structure of the network from a combination of missing information: partially reporting nodes and noisy activation times. We assume that no prior information on the topology is known to the algorithm. Due to many short loops, learning spreading parameters in the Zachary Club network using SLICER+ is not an easy task. As shown in Fig.~\ref{fig:zachary}(c), $10\%$ of unobserved nodes leads to a low error in parameter recovery of $5\%$ error on average. With $25\%$ of unobserved nodes, this error grows above $10\%$. However, the structure recovery is performed with a greater accuracy, despite no assumed knowledge on the edge set. Obviously, no algorithm can produce perfect reconstruction for high fractions of non-reporting nodes and unobserved edges due to a degeneracy in the space of solutions.

An explicit structure learning example produced for $\xi = 10\%$ unobserved nodes and $M = 1000$ cascades of length $T = 5$ is shown in Fig. \ref{fig:zachary}(a).
To eliminate possibility of bias due to a choice of ground-truth $\alpha_{ij}$, we use the test case with all $\alpha_{ij} = 0.5$.
The algorithm however does not have access to this information during the learning procedure.
Although the parameter reconstruction quality could be slightly improved by increasing the number of cascades, one can already see that the reconstruction on the observed part is perfect.
Incorrect classification appears only in the vicinity of the unobserved nodes.
Specifically, the problem is more pronounced for the two connected unobserved nodes, whereas the result is much better for an isolated unobserved node.

\begin{figure*}
    \includegraphics{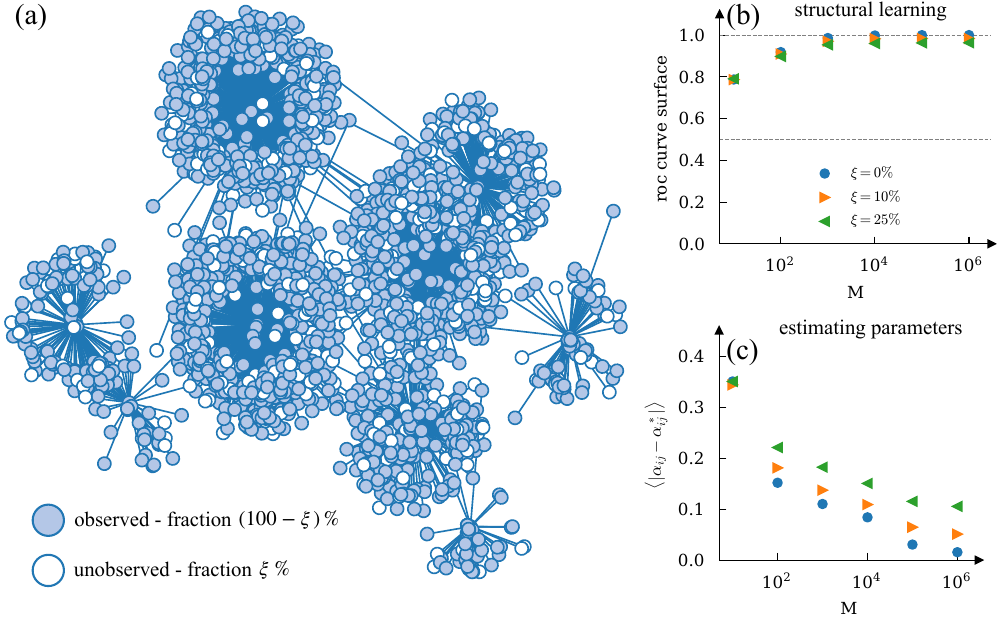}
    \caption{\label{fig:fb} (a) Structure of the FB network.
    (b) Average ROC curve surface, as a function of the number of available cascades for the Facebook network in the case where a fraction $\xi$ of nodes is unobserved.
    (c) Average difference between inferred $\alpha_{ij}$ and real $\alpha^*_{ij}$ parameters in $\ell_1$ norm, as a function of the number of available cascades for the Facebook network in the case where a fraction $\xi$ of nodes is unobserved.
    In both cases, each point is averaged over 5 different sets of parameters $\alpha^*_{ij}$ (sampled from a uniform distribution in $[0, 1]$).
    Network contains $N=2888$ nodes and $|E|=2981$ edges.
    Unobserved nodes were picked at random.
    All cascades had length $T=5$.
    We assume that a super-set of $5962$ edges is known, out of which we would like to infer the ground-truth set of $2981$ edges.
    Additionally, observed activation times are subject to known noise described in Eq. \ref{eq:synth_noise}.
    }
\end{figure*}

We now test structure recovery with SLICER+ on the larger Facebook network with the following setting: the observed timestamps are noisy, and we start with a superset of $5962$ edges, twice as large as the ground-truth set of $2981$ edges. The Facebook snapshot network is a sparse graph with multiple hubs. Hubs tend to increase the number of triangles, but the clustering of this network is on average lower than for the Zachary Club.
This is consistent with our observation that the reconstruction error in spreading parameters is significantly smaller, even when $25\%$ of nodes are unobserved, as shown in Fig. \ref{fig:fb}.
Furthermore, inclusion of the knowledge on the super-set of $5962$ potential edges allows to nearly perfectly reconstruct the network structure in the partial observation regime, assuming enough data is available.

\section{Conclusions}

Literature on learning of spreading models from data is rich, but not much attention so far was given to the problem of learning from incomplete and uncertain information.
Our work proposes a flexible method that addresses this problem.
SLICER+ deals with multiple settings related to incomplete and noisy data, that include partial observations in both spatial and temporal dimensions, noisy timestamps, unknown network structure, and combinations of these settings.

For unobserved times, we show that a proper formulation of the algorithm is able to learn model parameters from final and initial states only, assuming that enough data available.
This holds even in the presence of unobserved nodes.
We further show that uncertainty in provided activation times leads to significant bias, when not properly accounted for.

Although the structure learning task under unobserved nodes may generate degenerate solutions, we empirically showed that additional knowledge can significantly increase the quality of reconstruction using several representative synthetic networks, as well as two popular real-world networks. For real-world instances specifically, we used a setting where different types of incomplete and uncertain data settings are combined.

We found that our results are quite robust regardless of the synthetic network type. As an illustration, we considered quality of reconstruction as a function of the size of the set of unobserved nodes on a variety of random graph families.  

Our framework can be further developed in several future directions. In the setting of noisy observations, we assumed for simplicity that the noise distribution is known. However, the noise distribution probabilities $\pi_k$ can be treated as additional parameters and learned from the data. For the structure learning task, we considered reconstruction from a super-set of edges. It would be interesting to test the impact of other graph topological prior information on learning, such as details on density or degree distribution. Such a setting would be relevant in the setting where surveillance and data gathering procedures can be controlled. Under the scenario where the observations are too sparse and the degeneracy in the reconstruction is unavoidable and in applications focused on downstream prediction tasks, it could be useful to study the properties of the learned models as  \emph{effective} representations \cite{wilinski2021prediction}. Finally, in the future, it would be useful to extend our framework to other dynamic models, including models with reversible dynamics.\\

\section*{Code and Supplementary Material}

Full implementation of all algorithms studied in this work is available at \cite{wilinski2021code}.

\section*{Acknowledgments}

Authors acknowledge support from the Laboratory Directed Research and Development program of Los Alamos National Laboratory under project numbers 20240245ER and 20240198ER, and from U.S. DOE/SC Advanced Scientific Computing Research Program.

\onecolumngrid

\appendix

\section{Derivation of the DMP equations for the IC model}\label{app:bp}

The DMP equations presented in the paper are quite intuitive.
In case of simple models, such as IC model, they could be derived by identifying the correct dynamical variables to use in the equations, as it was done here.
They are also intrinsically connected to a more general framework of Belief Propagation (BP) equations, which are well described in detail in the literature \cite{mezard2009information, yedidia2003understanding}. In the dynamic setting, the BP equations on time trajectories have been studied by Kanoria and Montanari in \cite{kanoria2011majority}.
Here, for completeness, we show how DMP equations for IC model can be obtained directly from BP formulation.

Denote  $\Vec{\sigma}_i = (\sigma_i^0, \dots, \sigma_i^T)$ the time trajectory of node $i$ variable, and $w_i\left( \sigma_i^{t+1} | \,\sigma_i^t, \,\{ \sigma_j^t \}_{j \in \partial i} \right)$ the local probability of node $i$ transitioning to state $\sigma_i^{t+1}$ at time $t+1$, given its previous state and the states of its neighborhood at time $t$, and $P(\{ \sigma_i^0 \}_{i \in V})$ is the probability of the initial state. Additionally, let us introduce new type of message $\mu^{i \rightarrow j}(\Vec{\sigma}_i || \Vec{\sigma}_j)$, which represents the probability of node $i$ variable having the trajectory $\Vec{\sigma}_i$ on an auxiliary graph, where the trajectory of node $j$ is fixed to $\Vec{\sigma}_j$.
Our starting point is the dynamic belief propagation (DBP) equation on time trajectories \cite{kanoria2011majority}:
\begin{equation}
    \mu^{i \rightarrow j}_t(\Vec{\sigma}_i || \Vec{\sigma}_j) =
    \sum_{\mathclap{\{\sigma^0_k, \dots \sigma^{t-1}_k\}_{k \in \partial i \setminus j}}} P(\{ \sigma_i^0 \}_{i \in V}) \prod_{\tau=0}^{t-1} w_i\left( \sigma_i^{\tau+1} | \,\sigma_i^\tau, \,\{ \sigma_k^\tau \}_{k \in \partial i} \right) \prod_{\mathclap{k \in \partial i \setminus j}} \mu^{k \rightarrow i}_{t-1}(\Vec{\sigma}_k || \Vec{\sigma}_i),
\end{equation}
where $\mu^{i \rightarrow j}_t(\Vec{\sigma}_i || \Vec{\sigma}_j)$ is a message probability for trajectories up until time $t$.

A direct use of the above equations is still impractical due to their exponential complexity over cascade length $T$.
This is where the unidirectional character of the dynamics come in handy and leads to simplifications. In the case of the IC model, vector $\Vec{\sigma}_i$ can be fully described by a single number $\tau_i$, which represents the activation time.
This allows us to rewrite the DBP equations as follows:
\begin{equation}
    \mu^{i \rightarrow j}_{T+1}(\tau_i || \tau_j) =
    \sum_{\mathclap{\{\tau_k\}_{k \in \partial i \setminus j}}} \bar{W}_i(\tau_i; \{ \tau_k \}_{k \in \partial i}) \prod_{\mathclap{k \in \partial i \setminus j}} \mu_T^{k \rightarrow i}(\tau_k || \tau_i),
\end{equation}
\begin{equation}
    \mu^i_{T+1}(\tau_i) =
    \sum_{\mathclap{\{\tau_k\}_{k \in \partial i}}} \bar{W}_i(\tau_i; \{ \tau_k \}_{k \in \partial i}) \prod_{\mathclap{k \in \partial i}} \mu_T^{k \rightarrow i}(\tau_k || \tau_i),
\end{equation}
where $\mu^{i \rightarrow j}_T(\tau_i || \tau_j)$ and $\mu^i_T(\tau_i)$ correspond the same messages and marginals as before, but parameterized with activation times $\tau_i$.
Additionally, we denote $\bar{W}_i(\tau_i; \{ \tau_k \}_{k \in \partial i}) = P(\{ \sigma_i^0 \}_{i \in V}) \cdot W_i(\tau_i; \{ \tau_k \}_{k \in \partial i})$ for simplicity.
Before we move further we want to highlight some observations, which will be used later.
First, all the information required to compute the probability of activation at certain time $\tau_i$ is available at that time.
\begin{equation}
    \mu^i_T(\tau_i) = \mu^i_{\tau_i}(\tau_i) \quad \forall_{T > \tau_i},
\end{equation}
\begin{equation}
    \mu^{i \rightarrow j}_T(\tau_i || \tau_j) = \mu^{i \rightarrow j}_{\tau_i}(\tau_i || \tau_j) \quad \forall_{T > \tau_i}.
\end{equation}
Second, if the activation time $\tau_i$ of the node $i$ is earlier then the activation time $\tau_k$ of its neighbor $k$, than the message from $i$ to $k$ does not depend on $\tau_k$.
If we combine this we the previous equation, we can write:
\begin{equation}
    \mu^{i \rightarrow k}_T(\tau_i || \tau_k) = \mu^i_T(\tau_i || \infty) \quad \forall_{\tau_k > T},
\end{equation}
where $\infty$ describes the state where a given node was activated at any time later than the cutoff time $T$ (the time subscript).
Finally, the incoming messages in equations above are independent of node $i$ activation time $\tau_i$ and always behave as $\tau_i$ is later than their cutoff time:
\begin{equation}
    \mu^{i \rightarrow j}_T(\tau_i || \tau_j) =
    \sum_{\mathclap{\{\tau_k\}_{k \in \partial i \setminus j}}} \bar{W}_i(\tau_i; \{ \tau_k \}_{k \in \partial i}) \prod_{\mathclap{k \in \partial i \setminus j}} \mu_{T-1}^{k \rightarrow i}(\tau_k || \infty),
\end{equation}
\begin{equation}
    \mu^i_T(\tau_i) =
    \sum_{\mathclap{\{\tau_k\}_{k \in \partial i}}} \bar{W}_i(\tau_i; \{ \tau_k \}_{k \in \partial i}) \prod_{\mathclap{k \in \partial i}} \mu_{T-1}^{k \rightarrow i}(\tau_k || \infty).
\end{equation}
This is a direct consequence of the previous observations.
Let us now define new type of marginals and messages:
\begin{equation}
    P_S^i(t) = \sum_{\tau_i > t} \mu_T^i(\tau_i) = \mu_t^i(*),
\end{equation}
\begin{equation}
    P_S^{i \rightarrow j}(t) = \sum_{\tau_i > t} \mu_T^{i \rightarrow j}(\tau_i || \infty) = \mu_t^{i \rightarrow j}(* || \infty),
\end{equation}
where $P_S^i(t)$ is the probability that node $i$ was not activated until time $t$ and $P_S^{i \rightarrow j}(t)$ is the same probability, but on an auxiliary graph where node $j$ does not exist.
Symbol $*$ stands for all the trajectories with the activation time $\tau_i$ being after the time horizon $T$.
It means that when we sum over $\tau_i > b$, in practice we sum over $\tau_i \in \{b+1, b+2, \dots, T-1, T, *\}$.
It should be noted that in all equations below we use the convention that when $\tau_i = *$, $\tau_i - 1 = T$.
Now, we use these new objects to compute the term $\bar{W}_i(\dots)$, which describes the dynamics of the process.
In the case of IC model it has the following form:
\begin{equation}
    \begin{split}
        \bar{W}_i(\tau_i; \{ \tau_k \}_{k \in \partial i}) &= (1 - P_S^i(0)) \cdot \mathds{1}(\tau_i = 0)\\
        &+ P_S^i(0) \cdot \mathds{1}(\tau_i > 0) \left[\prod_{k \in \partial i} 1 - \alpha_{ki} \cdot \mathds{1}(\tau_k < \tau_i - 1) \right]\\
        &\times \left[1 - \prod_{k \in \partial i} (1 - \alpha_{ki} \cdot \mathds{1}(\tau_k = \tau_i - 1)) \cdot \mathds{1}(\tau_i \neq *) \right]
    \end{split}
\end{equation}
We can now plug it all together and get:
\begin{equation}
    \begin{split}
        1 - P_S^{i \rightarrow j}(t) &= 1 -
        \sum_{\mathclap{\{\tau_k\}_{k \in \partial i \setminus j}}} \bar{W}_i(*; \{ \tau_k \}_{k \in \partial i}) \prod_{\mathclap{k \in \partial i \setminus j}} \mu_{t-1}^{k \rightarrow i}(\tau_k || \infty)\\
        &= 1 - P_S^i(0) \sum_{\{\tau_k\}_{k \in \partial i \setminus j}} \prod_{k \in \partial i} \left( 1 - \alpha_{ki} \cdot \mathds{1}(\tau_k \neq *) \right) \prod_{\mathclap{k \in \partial i \setminus j}} \mu_{t-1}^{k \rightarrow i}(\tau_k || \infty)\\
        &= 1 - P_S^i(0) \sum_{\{\tau_k\}_{k \in \partial i \setminus j}} \prod_{k \in \partial i \setminus j} \left( 1 - \alpha_{ki} \cdot \mathds{1}(\tau_k \neq *) \right) \mu_{t-1}^{k \rightarrow i}(\tau_k || \infty)\\
        &= 1 - P_S^i(0) \prod_{k \in \partial i \setminus j} \sum_{\tau_k} \left( 1 - \alpha_{ki} \cdot \mathds{1}(\tau_k \neq *) \right) \mu_{t-1}^{k \rightarrow i}(\tau_k || \infty)\\
        &= 1 - P_S^i(0) \prod_{k \in \partial i \setminus j} \left( \sum_{\tau_k \leq t-1} (1 - \alpha_{ki}) \mu_{t-1}^{k \rightarrow i}(\tau_k || \infty) + \mu_{t-1}^{k \rightarrow i}(* || \infty) \right)\\
        &= 1 - P_S^i(0) \prod_{k \in \partial i \setminus j} \bigg(1 - \alpha_{ki} (1 - P_S^{k \rightarrow i}(t-1))\bigg),
    \end{split}
\end{equation}
which is equivalent to the DMP equations (\ref{eq:message}) from the main text, where $p_{i \rightarrow j}(t) =  1 - P_S^{i \rightarrow j}(t)$ and assuming some initial condition $\bar{p}_{i \rightarrow j} =  1 - P_S^{i \rightarrow j}(0)$.
Readers interested in derivation of DMP from DBP for different dynamic models should refer to \cite{kanoria2011majority, lokhov2015dynamic, sun2021competition}.

\section{Derivation of SLICER+ for different scenarios of incomplete data}
\label{app:SLICER+_derivation}

Below we present the details of Lagrangian formulation used in the optimisation scheme of each of the different case of missing or incomplete data.
We show step-by-step how to compute all the Lagrange multipliers and the gradient step used to update the parameters.

\subsection{SLICER}\label{app:slicer_form}

The SLICER algorithm \cite{wilinski2021prediction} was developed to deal with the scenario of missing information on nodes exclusively.
In this setting the objective takes the following form:
\begin{equation}
    \mathcal{O} = \sum_{s \in S} \sum_{i \in O} \sum_{\tau_i^s} m^{\tau^s_i} \log \left( p^s_i(t) \cdot \mathds{1}_{(t \leq T)} - p^s_i(t - 1) \cdot \mathds{1}_{(t > 0)} + \mathds{1}_{(t = *)} \right).
\end{equation}
The constraints are the same as in the main text, but lets remind them for reader's convenience.
\begin{equation}
    \begin{split}
        \mathcal{C} &= \sum_{s \in S} \sum_{t=0}^{T-1} \sum_{i \in V} \lambda^s_i(t+1) \Bigg( p^s_i(t+1) - 1 + \big( 1 - \bar{p}^s_i \big) \prod_{k \in \partial i} \bigg(1 - \alpha_{ki} \cdot p^s_{k \rightarrow i}(t) \bigg) \Bigg) \\
        &+\sum_{s \in S} \sum_{t=0}^{T-1} \sum_{(i,j) \in E} \lambda^s_{i \rightarrow j}(t+1) \Bigg( p^s_{i \rightarrow j}(t+1) - 1 + \big( 1 - \bar{p}^s_i \big) \prod_{k \in \partial i \setminus j} \bigg(1 - \alpha_{ki} \cdot p^s_{k \rightarrow i}(t) \bigg) \Bigg).
    \end{split}
\end{equation}
The sum of the objective and the constraints constitutes the Lagrangian.
Lagrange multipliers can now be found by differentiating the Lagrangian:
\begin{equation}
    \begin{split}
        \frac{\partial \mathcal{L}}{\partial p^s_i(t)} &= \lambda^s_i(t) + \sum_{\tau_i^s} \frac{m^{\tau^s_i} \cdot \mathds{1}_{(t=\tau^s_i)} \cdot \mathds{1}_{(\tau^s_i \leq T)}}{p^s_i(\tau^s_i) - p^s_i(\tau^s_i - 1) \cdot \mathds{1}_{(\tau^s_i > 0)}} \\
        &+ \sum_{\tau_i^s} \frac{m^{\tau^s_i} \cdot \mathds{1}_{(t=\tau^s_i-1)} \cdot \mathds{1}_{(\tau^s_i > 0)}}{p^s_i(\tau^s_i - 1) - p^s_i(\tau^s_i) \cdot \mathds{1}_{(\tau^s_i \leq T)} - \mathds{1}_{(\tau^s_i = *)}},
    \end{split}
    \label{eq:lagrange_marginal}
\end{equation}
\begin{equation}
    \begin{split}
        \frac{\partial \mathcal{L}}{\partial p^s_{i \rightarrow j}(t)} &= \lambda^s_{i \rightarrow j}(t) - \lambda^s_j(t+1) \, \alpha_{ij} \, \big( 1 - \bar{p}^s_j \big) \prod_{m \in \partial j \setminus i} \Big(1 - \alpha_{mj} \cdot p^s_{m \rightarrow j}(t) \Big) \\
        &- \sum_{k \in \partial j \setminus i} \lambda^s_{j \rightarrow k}(t+1) \, \alpha_{ij} \, \big( 1 - \bar{p}^s_j \big) \prod_{m \in \partial j \setminus \{ i, k \}} \Big(1 - \alpha_{mj} \cdot p^s_{m \rightarrow j}(t) \Big)
    \end{split}
    \label{eq:lagrange_message}
\end{equation}
and equating the derivatives to zero.
Finally, derivatives over parameters $\alpha_{ij}$ read
\begin{equation}
    \begin{split}
        \frac{\partial \mathcal{L}}{\partial \alpha_{ij}} = &- \sum_{s \in S} \sum_{t=0}^{T-1} \lambda^s_i(t+1) \, p^s_{j \rightarrow i}(t) \, \big( 1 - \bar{p}^s_i \big) \prod_{m \in \partial i \setminus j} \Big(1 - \alpha_{mi} \cdot p^s_{m \rightarrow i}(t) \Big) \\
        &-\sum_{s \in S} \sum_{t=0}^{T-1} \lambda^s_j(t+1) \, p^s_{i \rightarrow j}(t) \, \big( 1 - \bar{p}^s_j \big) \prod_{m \in \partial j \setminus i} \Big(1 - \alpha_{mj} \cdot p^s_{m \rightarrow j}(t) \Big) \\
        &-\sum_{s \in S} \sum_{t=0}^{T-1} \sum_{k \in \partial i \setminus j} \lambda^s_{i \rightarrow k}(t+1) \, p^s_{j \rightarrow i}(t) \, \big( 1 - \bar{p}^s_i \big) \prod_{m \in \partial i \setminus \{ j, k \}} \Big(1 - \alpha_{mi} \cdot p^s_{m \rightarrow i}(t) \Big) \\
        &-\sum_{s \in S} \sum_{t=0}^{T-1} \sum_{k \in \partial j \setminus i} \lambda^s_{j \rightarrow k}(t+1) \, p^s_{i \rightarrow j}(t) \, \big( 1 - \bar{p}^s_j \big) \prod_{m \in \partial j \setminus \{ i, k \}} \Big(1 - \alpha_{mj} \cdot p^s_{m \rightarrow j}(t) \Big),
    \end{split}
\end{equation}
which can further be simplified for $\alpha_{ij} \neq 0$:
\begin{equation}
    \frac{\partial \mathcal{L}}{\partial \alpha_{ij}} = -\frac{1}{\alpha_{ij}} \sum_{s \in S} \sum_{t=0}^{T-1} \left(\lambda^s_{i \rightarrow j}(t) \, p^s_{i \rightarrow j}(t) + \lambda^s_{j \rightarrow i}(t) \, p^s_{j \rightarrow i}(t)\right).
    \label{eq:alpha}
\end{equation}
The above can be directly used in the forward–backward propagation procedure described in the main text.  

\subsection{Simple Graphs}\label{app:simple_form}

In the case of simple graphs, where $\alpha_{ij} = \alpha \, \forall_{(i,j) \in E}$, the objective is the same as for the original SLICER, but the constraints simplify to:
\begin{equation}
    \begin{split}
        \mathcal{C} &= \sum_{s \in S} \sum_{t=0}^{T-1} \sum_{i \in V} \lambda^s_i(t+1) \Bigg( p^s_i(t+1) - 1 + \big( 1 - \bar{p}^s_i \big) \prod_{k \in \partial i} \bigg(1 - \alpha \cdot p^s_{k \rightarrow i}(t) \bigg) \Bigg) \\
        &+\sum_{s \in S} \sum_{t=0}^{T-1} \sum_{(i,j) \in E} \lambda^s_{i \rightarrow j}(t+1) \Bigg( p^s_{i \rightarrow j}(t+1) - 1 + \big( 1 - \bar{p}^s_i \big) \prod_{k \in \partial i \setminus j} \bigg(1 - \alpha \cdot p^s_{k \rightarrow i}(t) \bigg) \Bigg).
    \end{split}
\end{equation}
As a result, we need to update the derivatives over messages:
\begin{equation}
    \begin{split}
        \frac{\partial \mathcal{L}}{\partial p^s_{i \rightarrow j}(t)} &= \lambda^s_{i \rightarrow j}(t) - \lambda^s_j(t+1) \, \alpha \, \big( 1 - \bar{p}^s_j \big) \prod_{m \in \partial j \setminus i} \Big(1 - \alpha \cdot p^s_{m \rightarrow j}(t) \Big) \\
        &- \sum_{k \in \partial j \setminus i} \lambda^s_{j \rightarrow k}(t+1) \, \alpha \, \big( 1 - \bar{p}^s_j \big) \prod_{m \in \partial j \setminus \{ i, k \}} \Big(1 - \alpha \cdot p^s_{m \rightarrow j}(t) \Big),
    \end{split}
\end{equation}
while the derivatives over marginals remain in the same form as before.
Then we can rewrite the derivatives over parameters, which in this case reduce to only a single equation:
\begin{equation}
    \begin{split}
        \frac{\partial \mathcal{L}}{\partial \alpha} = &- \sum_{s \in S} \sum_{t=0}^{T-1} \sum_{i \in V} \lambda^s_i(t+1) \sum_{j \in \partial i} p^s_{j \rightarrow i}(t) \, \big( 1 - \bar{p}^s_i \big) \prod_{m \in \partial i \setminus j} \Big(1 - \alpha \cdot p^s_{m \rightarrow i}(t) \Big) \\
        &-\sum_{s \in S} \sum_{t=0}^{T-1} \sum_{(i, k) \in E'} \lambda^s_{i \rightarrow k}(t+1) \sum_{j \in \partial i \setminus k} p^s_{j \rightarrow i}(t) \, \big( 1 - \bar{p}^s_i \big) \prod_{m \in \partial i \setminus \{ j, k \}} \Big(1 - \alpha \cdot p^s_{m \rightarrow i}(t) \Big),
    \end{split}
\end{equation}
where by $E'$ we denote the set of edges such that $(i, j)$ and $(j, i)$ are counted as separate elements.
The above equation can, assuming that $\alpha > 0$, be simplified to:
\begin{equation}
    \frac{\partial \mathcal{L}}{\partial \alpha} = -\frac{1}{\alpha} \sum_{s \in S} \sum_{t=0}^{T-1} \sum_{(i, j) \in E'} \lambda^s_{i \rightarrow j}(t) \cdot p^s_{i \rightarrow j}(t).
\end{equation}
Now we can repeat the whole procedure, but the update step is in only one dimension.

\subsection{Missing Times}\label{app:times_form}

In this setting part of the activation times are known precisely, while others are known to be in certain time intervals.
Based on that we modify the objective in a following way.
\begin{equation}
    \begin{split}
        \mathcal{O} &= \sum_{s \in S} \sum_{i \in O} \sum_{\tau^s_i} m^{\tau^s_i} \log\Big( p^s_i(\tau^s_i) \cdot \mathds{1}_{(\tau^s_i \leq T)} - p^s_i(\tau^s_i - 1) \cdot \mathds{1}_{(\tau^s_i > 0)} + \mathds{1}_{(\tau^s_i = *)} \Big) \\
        &+ \sum_{s \in S} \sum_{i \in O} \sum_{\hat{\tau}^s_i} m^{\hat{\tau}^s_i} \log\Big( p^s_i(\hat{\tau}^s_i) - p^s_i(\hat{\tau}^s_i - \Delta\hat{\tau}^s_i) \Big).
    \end{split}
\end{equation}
The only change is that if the node was activated during an unobserved time interval, the marginal probability of being infected precisely at $\tau_i$ is replaced by a marginal probability of being infected inside the interval $[\hat{\tau}_i - \Delta \hat{\tau}_i, \hat{\tau}_i]$.
The constraints are the same as in the original SLICER.
The above modification affects only the computation of the Lagrange derivative over marginal probabilities.
\begin{equation}
    \begin{split}
        \frac{\partial \mathcal{L}}{\partial p^s_i(t)} &= \lambda^s_i(t) + \sum_{\tau_i^s} \frac{m^{\tau^s_i} \cdot \mathds{1}_{(t=\tau^s_i)} \cdot \mathds{1}_{(\tau^s_i \leq T)}}{p^s_i(\tau^s_i) - p^s_i(\tau^s_i - 1) \cdot \mathds{1}_{(\tau^s_i > 0)}} \\
        &+ \sum_{\tau_i^s} \frac{m^{\tau^s_i} \cdot \mathds{1}_{(t=\tau^s_i-1)} \cdot \mathds{1}_{(\tau^s_i > 0)}}{p^s_i(\tau^s_i - 1) - p^s_i(\tau^s_i) \cdot \mathds{1}_{(\tau^s_i \leq T)} - \mathds{1}_{(\tau^s_i = *)}} \\
        &+ \sum_{\hat{\tau}_i^s} \frac{m^{\hat{\tau}^s_i} \cdot \mathds{1}_{(t=\hat{\tau}^s_i)}}{p^s_i(\hat{\tau}^s_i) - p^s_i(\hat{\tau}^s_i - \Delta\hat{\tau}^s_i)} + \sum_{\hat{\tau}_i^s} \frac{m^{\hat{\tau}^s_i} \cdot \mathds{1}_{(t=\hat{\tau}^s_i-\Delta\hat{\tau}^s_i)}}{p^s_i(\hat{\tau}^s_i - \Delta\hat{\tau}^s_i) - p^s_i(\hat{\tau}^s_i)}.
    \end{split}
\end{equation}
This requires updating the values of Lagrange multipliers $\lambda^s_i(t)$.
The rest of the procedure remains the same.

\subsection{Noisy Timestamps}\label{app:noisy_form}

Taking into account the noisy activation times results with an objective function of the following form:
\begin{equation}
    \mathcal{O} = \sum_{s \in S} \sum_{i \in O} \sum_{\tau^s_i} m^{\tau^s_i} \log\left( \sum_{k=-K}^K \pi_k \bigg[ p^s_i(\tau^s_i - k) \cdot \mathds{1}_{(0 \leq \tau^s_i - k \leq T)} - p^s_i(\tau^s_i - 1 - k) \cdot \mathds{1}_{(0 < \tau^s_i - k \leq *)} + \mathds{1}_{(\tau^s_i - k = *)} \bigg]\right).
\end{equation}
Updating the objective function affects the learning procedure.
However, the only element of the optimisation scheme, which require changes is the computation of $\lambda_i^s(t)$.
\begin{equation}
    \frac{\partial \mathcal{L}}{\partial p^s_i(t)} = \lambda^s_i(t) + \sum_{\tau_i^s} \frac{m^{\tau^s_i} \cdot \sum_{k=-K}^K \pi_k \left( \mathds{1}_{(t=\tau^s_i-k)} \cdot \mathds{1}_{(0 \leq \tau^s_i-k \leq T)} - \mathds{1}_{(t=\tau^s_i-1-k)} \cdot \mathds{1}_{(0 < \tau^s_i-k \leq *)} \right)}{\sum_{k=-K}^K \pi_k \left( p^s_i(\tau^s_i - k) \cdot \mathds{1}_{(0 \leq \tau^s_i - k \leq T)} - p^s_i(\tau^s_i - 1 - k) \cdot \mathds{1}_{(0 < \tau^s_i - k \leq *)} + \mathds{1}_{(\tau^s_i - k = *)} \right)}.
\end{equation}
All the further computations remain the same as before.
It should be noted that the described change in the learning procedure does also affect its complexity.
In the worst case scenario where $K=T$ the complexity of the algorithm becomes $\max[O(|E| |S| T), O(N |S| T^2)]$ in comparison with the complexity $O(|E| |S| T)$ of the original algorithm.

\section{Efficient Implementation}\label{app:eff}

Direct implementation of Eq. (\ref{eq:marginal}) and (\ref{eq:message}) suffers from evaluating the products over neighbors, in particular, this is repeated for every message rooted in a given node.
As a result the complexity of the algorithm depends on the network's degree distribution, instead of the number of edges.
Networks with broad degree distributions, which are common in real-world applications, are specifically affected by this problem.
It is possible to overcome this by rewriting the DMP equations in the following form:
\begin{equation}
    p^c_i(t) = 1 - \big( 1 - \bar{p}^c_i \big) \prod_{k \in \partial i} \Big(1 - \alpha_{ki} \cdot p^c_{k \rightarrow i}(t - 1) \Big),
    \label{eq:eff_marginal}
\end{equation}
\begin{equation}
    p^c_{j \rightarrow i}(t) = 1 - \frac{1 - p^c_j(t)}{1 - \alpha_{ij} \cdot p^c_{i \rightarrow j}(t - 1)},
    \label{eq:eff_message}
\end{equation}
where the marginals equations are the same, but the messages are computed using the pre-computed marginals.
Now the complexity is independent of the degree distribution and truly linear in the number of edges, assuming that $\alpha_{ij} \cdot p^c_{i \rightarrow j}(t - 1) \neq 1$.
The latter is, however, a reasonable assumption, which is unlikely to be broken, in which case the original equation can still be used.
Similar problem arises from Eq. (\ref{eq:lagrange_message}), which after equating left hand side to zero leads to:
\begin{equation}
    \begin{split}
        \lambda^s_{i \rightarrow j}(t) &= \lambda^s_j(t+1) \, \alpha_{ij} \, \big( 1 - \bar{p}^s_j \big) \prod_{m \in \partial j \setminus i} \Big(1 - \alpha_{mj} \cdot p^s_{m \rightarrow j}(t) \Big) \\
        &+ \sum_{k \in \partial j \setminus i} \lambda^s_{j \rightarrow k}(t+1) \, \alpha_{ij} \, \big( 1 - \bar{p}^s_j \big) \prod_{m \in \partial j \setminus \{ i, k \}} \Big(1 - \alpha_{mj} \cdot p^s_{m \rightarrow j}(t) \Big).
    \end{split}
\end{equation}
Although the equation have a different form, they are dual to the DMP one.
Similarly to the latter, direct implementation leads to degree distribution dependence.
Here instead of a product over neighbors we have a sum over neighbors and similarly as before, this can be prevented.
To do so, let us introduce
\begin{equation}
    \widehat{\lambda}^s_j(t) = \sum_{k \in \partial j} \lambda^s_{j \rightarrow k}(t+1) \, \big( 1 - \bar{p}^s_j \big) \prod_{m \in \partial j \setminus k} \Big(1 - \alpha_{mj} \cdot p^s_{m \rightarrow j}(t) \Big),
\end{equation}
which can be further simplified as
\begin{equation}
    \widehat{\lambda}^s_j(t) = \sum_{k \in \partial j} \lambda^s_{j \rightarrow k}(t+1) \, \big( 1 - p^s_{j \rightarrow k}(t+1) \big).
\end{equation}
Now we can write:
\begin{equation}\label{eq:smart}
    \begin{split}
        \lambda^s_{i \rightarrow j}(t) &= \lambda^s_j(t+1) \, \alpha_{ij} \, \big(1 - p^s_{j \rightarrow i}(t+1) \big) \\
        &+ \alpha_{ij} \, \frac{\widehat{\lambda}^s_j(t) - \lambda^s_{j \rightarrow i}(t+1) \, \big( 1 - p^s_{j \rightarrow i}(t+1) \big)}{1 - \alpha_{ij} \cdot p^s_{i \rightarrow j}(t)}.
    \end{split}
\end{equation}
This way, knowing that $\lambda^s_{i \rightarrow j}(T) = 0 \, \, \forall_{i, j}$, we reduced the complexity to being linear in the number of edges, same way as we did for messages.
Additionally, in the directed case where $(i, j) \in E$, but $(j, i) \notin E$, one needs to rewrite Eq. (\ref{eq:smart}) into:
\begin{equation}
    \lambda^s_{i \rightarrow j}(t) = \alpha_{ij} \, \frac{\widehat{\lambda}^s_j(t) + \lambda^s_j(t+1) \, \big(1 - p^s_j(t+1) \big)}{1 - \alpha_{ij} \cdot p^s_{i \rightarrow j}(t)}.
\end{equation}
The above equations, used in all our experiments, are implemented in Julia \cite{bezanson2011julia} and available at \cite{wilinski2021code}.

\section{Structure learning without any prior knowledge on the topology}\label{app:topo}

\begin{figure*}
    \includegraphics{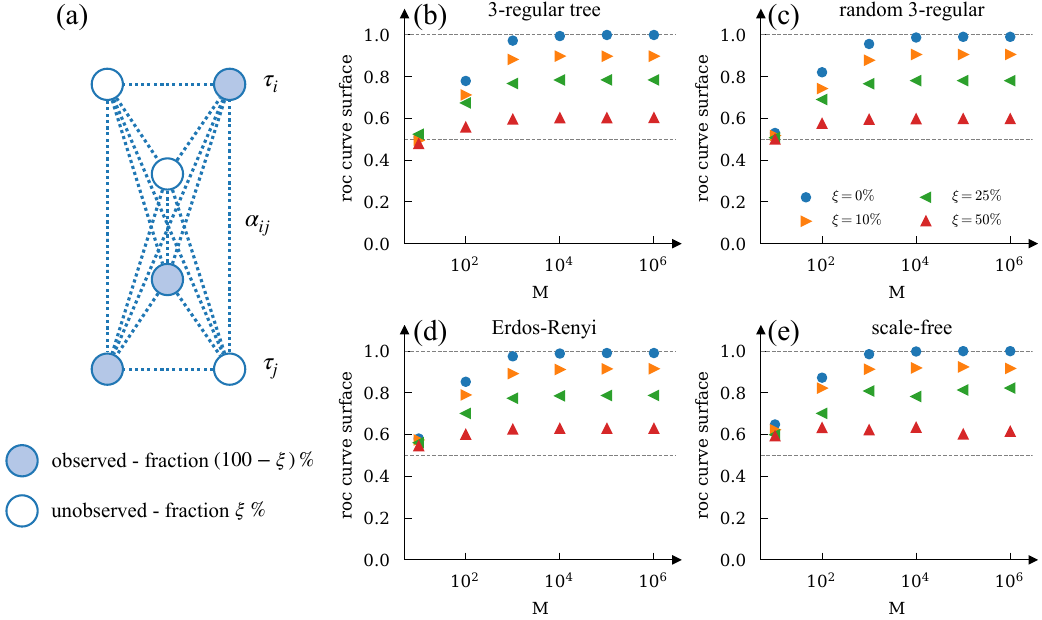}
    \caption{\label{fig:no_struct} Structure learning starting with a fully-connected graph. Average ROC curve surface, as a function of the number of available cascades for different network types in the case where a fraction $\xi$ of nodes is unobserved and there is no prior knowledge about the topology.
    Each point is averaged over 5 different networks and 5 different sets of parameters $\alpha^*_{ij}$ (sampled from a uniform distribution in $[0,1]$).
    All networks contain $N=100$ nodes and all but the tree have average degree equal to $\langle k \rangle = 3$.
    Unobserved nodes were picked at random.
    All cascades had length $T=5$.}
\end{figure*}

Here, we present results on structure learning in the presence of unobserved nodes. We reproduce the experiment presented in Fig.~\ref{fig:struct}, but in the most challenging setting where no information on the network topology is available, and the super-set of edges that we initially consider corresponds to a fully-connected graph.
The results are given in Fig.~\ref{fig:no_struct}, and are similar for all four synthetic network types.The edge reconstruction is always better than random guess, but we observe reduction of reconstruction quality with the increase of the number of unobserved nodes.
This is expected due to the appearance of degeneracy where several models are consistent with the observation data.
As a result SLICER produces a graph, which is compatible with the observed data and as such can be used as an effective model for prediction \cite{wilinski2021prediction}, even if it is not the \textit{true} graph.
Moreover, have in mind that our method recovers the network simultaneously with learning model parameters.

\twocolumngrid

\bibliographystyle{unsrt}
\bibliography{literature}

\end{document}